\documentclass[prd,showpacs,floatfix]{revtex4}
\usepackage{amsmath,amssymb,graphicx}

\begin{document}

\title{Stability of the Einstein static universe in IR modified
Ho\v{r}ava gravity}

\author{Christian G. B\"ohmer}
\email{c.boehmer@ucl.ac.uk}
\affiliation{Department of Mathematics and Institute of Origins,
University College London, Gower Street, London, WC1E 6BT, UK}

\author{Francisco S. N. Lobo}
\email{flobo@cii.fc.ul.pt} \affiliation{Centro de F\'{i}sica
Te\'{o}rica e Computacional, Faculdade de Ci\^{e}ncias da
Universidade de Lisboa, Avenida Professor Gama Pinto 2, P-1649-003
Lisboa, Portugal}

\date{\today}

\begin{abstract}

Recently, Ho\v{r}ava proposed a power counting renormalizable
theory for (3+1)-dimensional quantum gravity, which reduces to
Einstein gravity with a non-vanishing cosmological constant in IR,
but possesses improved UV behaviors. In this work, we analyze the
stability of the Einstein static universe by considering linear
homogeneous perturbations in the context of an IR modification of
Ho\v{r}ava gravity, which implies a `soft' breaking of the
`detailed balance' condition. The stability regions of the
Einstein static universe is parameterized by the linear equation
of state parameter $w=p/\rho$ and the parameters appearing in the
Ho\v{r}ava theory, and it is shown that a large class of stable
solutions exists in the respective parameter space.

\end{abstract}

\pacs{04.50.+h, 04.20.Jb, 04.25.Nx}
\maketitle

\section{Introduction}

Recently, motivated by the Lifshitz model in condensed matter
physics, Ho\v{r}ava proposed a power counting renormalizable
theory for (3+1)-dimensional quantum gravity
\cite{Horava:2008ih,Horava:2009uw}. This theory, denoted as
Ho\v{r}ava-Lifshitz gravity, is believed to be the potential
ultraviolet (UV) completion of general relativity (GR). In the
infrared (IR) limit (setting the parameter $\lambda= 1$ in the
action), it recovers GR. Ho\v{r}ava-Lifshitz gravity admits a
Lifshitz scale-invariance in time and space, exhibiting a broken
Lorentz symmetry at short scales, while at large distances higher
derivative terms do not contribute, and the theory reduces to
standard GR. Since then various properties and characteristics of
the Ho\v{r}ava gravities have been extensively analyzed, ranging
from formal developments \cite{formal}, cosmology
\cite{cosmology}, dark energy \cite{darkenergy,Park:2009zr} and
dark matter \cite{darkmatter}, spherically symmetric solutions
\cite{BHsolutions,Park:2009zra}, and its viability with
observational constraints \cite{HLtests} were also explored.
Although a generic vacuum of the theory is the anti-de Sitter one,
particular limits of the theory allow for the Minkowski vacuum. In
this limit post-Newtonian coefficients coincide with those of pure
GR. Thus, the deviations from conventional GR can be tested only
beyond the post-Newtonian corrections, that is for a system with
strong gravity at astrophysical scales.

In this work, we consider the stability of the Einstein static
universe in Ho\v{r}ava-Lifshitz gravity, with a `soft' violation
of the detailed balance condition (Recently, the stability of the
Einstein static universe in Ho\v{r}ava-Lifshitz gravity satisfying
the detailed balance condition was analyzed \cite{Wu:2009ah}). The
presence of the respective term in the action which represents a
`soft' violation of the `detailed balance' condition modifies the
IR behavior. Note that this IR modification term, with an
arbitrary cosmological constant, represent the analogs of the
standard Schwarzschild--(A)dS solutions, which were absent in the
original Ho\v{r}ava model. The analysis of the static Einstein
Universe is motivated by the possibility that the universe might
have started out in an asymptotically Einstein static state, in
the inflationary universe context \cite{Ellis:2002we}. On the
other hand, the Einstein cosmos has always been of great interest
in various gravitational theories.

In GR for instance, generalizations with non-constant pressure
have been analyzed in \cite{ESa}. In brane world models, the
Einstein static universe was investigated in
\cite{Gergely:2001tn}, while its generalization within
Einstein-Cartan theory can be found in \cite{Boehmer:2003iv}, and
in loop quantum cosmology, we refer the reader to
\cite{Mulryne:2005ef}. In the context of $f(R)$ modified theories
of gravity, the stability of the Einstein static universe was also
analyzed by considering homogeneous perturbations
\cite{Boehmer:2007tr}. By considering specific forms of $f(R)$,
the stability regions of the solutions were parameterized by a
linear equation of state parameter $w=p/\rho$. Contrary to
classical GR, it was found that in $f(R)$ gravity a stable
Einstein cosmos with a positive cosmological constant does indeed
exist. Thus, in principle, modifications in $f(R)$ gravity
stabilize solutions which are unstable in GR. Furthermore, in
\cite{Goswami:2008fs} it was found that only one class of $f(R)$
theories admits an Einstein static model, and that this class is
neutrally stable with respect to vector and tensor perturbations
for all equations of state on all scales. These results are
apparently contradictory with those of Ref.~\cite{Boehmer:2007tr}.
However, in a recent work, homogeneous and inhomogeneous scalar
perturbations in the Einstein static solutions were analyzed
\cite{Seahra:2009ft}, consequently reconciling both of the above
works. In the context of modified theories of gravity, the
stability of the Einstein static universe in $f(G)$ Gauss-Bonnet
modified gravity was also analyzed \cite{Bohmer:2009fc}. In
particular, by considering a generic form of $f(G)$, the stability
regions of the Einstein static universe were parameterized by the
linear equation of state and the second derivative $f''(G)$ of the
Gauss-Bonnet term. It was shown that stable modes for all equation
of state parameters $w$ exist, if the parameters of the theory are
chosen appropriately. Thus, the results show that perturbation
theory of modified theories of gravity present a richer
stability/instability structure than in GR.

Thus, it is the purpose of the present paper to consider the
stability of the Einstein static universe by considering linear
homogeneous perturbations in Ho\v{r}ava-Lifshitz gravity. Indeed,
this analysis is particularly important as the higher derivative
terms in the action contributes with a $1/a^4$ term in the
modified Friedman equations. This contribution becomes dominant
for small $a$, and as mentioned above motivates this analysis, due
to the possibility that the universe might have started out in an
asymptotically Einstein static state, in the inflationary universe
context \cite{Ellis:2002we}. On the other hand, the cosmological
solutions of GR are recovered at large scales. It is shown that a
large class of Einstein static universes exist that are stable
with respect to linear homogeneous perturbations.

This paper is outlined in the following manner: In Sec.
\ref{sec:II}, we briefly review the action and field equations of
Ho\v{r}ava gravity, and the respective modified Friedman
equations. In Sec. \ref{sec:III}, we consider linear homogeneous
perturbations in the context of the Einstein static Universe in
Ho\v{r}ava gravity, and analyze the respective stability regions.
In Sec. \ref{sec:concl}, we conlcude.

\section{Ho\v{r}ava gravity and field equations}
\label{sec:II}

\subsection{Action}

Using the ADM formalism, the four-dimensional metric is
parameterized by the following
\begin{equation}
ds^2=-N^2c^2\,dt^2+g_{ij}\left(dx^i+N^i\,dt\right)
\left(dx^j+N^j\,dt\right)\,,
\end{equation}
where $N$ is the lapse function, $N^i$ is the shift vector, and
$g_{ij}$ is the 3-dimensional spatial metric.

In this context, the Einstein-Hilbert action is given by
\begin{equation}
S=\frac{1}{16\pi G}\int d^4x \;\sqrt{g}\,N\left(K_{ij}K^{ij}-
K^2+R^{(3)}-2\Lambda \right) , \label{EHaction}
\end{equation}
where $G$ is Newton's constant, $R^{(3)}$ is the three-dimensional
curvature scalar for $g_{ij}$, and $K_{ij}$ is the extrinsic
curvature defined as
\begin{equation}
K_{ij}=\frac{1}{2N}\left(\dot{g}_{ij}-\nabla_iN_j-\nabla_jN_i\right),
\end{equation}
where the overdot denotes a derivative with respect to $t$, and
$\nabla_i$ is the covariant derivative with respect to the spatial
metric $g_{ij}$.

Consider the IR-modified Ho\v{r}ava action given by
\begin{eqnarray}
S&=&\int dt\,d^3t
\;\sqrt{g}\,N\Bigg[\frac{2}{\kappa^2}\left(K_{ij}K^{ij}-\lambda
K^2\right) -\frac{\kappa^2}{2\nu^4}
C_{ij}C^{ij}+\frac{\kappa^2\mu}{2\nu^2}\epsilon^{ijk}R^{(3)}_{il}\nabla_j
R^{(3)l}{}_{k}
     \nonumber  \\
&&-\frac{\kappa^2\mu^2}{8}R^{(3)}_{ij}R^{(3)ij}+\frac{\kappa^2\mu^2}
{8(3\lambda-1)} \left(\frac{4\lambda-1}{4}(R^{(3)})^2-\Lambda_W
R^{(3)}+3\Lambda_W^2\right)+\frac{\kappa^2\mu^2\varpi}{8(3\lambda-1)}
R^{(3)}\Bigg]\,, \label{Haction}
\end{eqnarray}
where $\kappa$, $\lambda$, $\nu$, $\mu$, $\varpi$ and $\Lambda_W$
are constant parameters. $C^{ij}$ is the Cotton tensor, defined as
\begin{equation}
C^{ij}=\epsilon^{ikl}\nabla_k\left(R^{(3)j}{}_{l}-\frac{1}{4}
R^{(3)}\delta^j_{l}\right)\,.
\end{equation}
Note that the last term in Eq. (\ref{Haction}) represents a `soft'
violation of the `detailed balance' condition, which modifies the
IR behavior. This IR modification term, $\mu^4 R^{(3)}$,
generalizes the original Ho\v{r}ava model (we have used the
notation of Ref. \cite{Park:2009zra}). Note that now these
solutions with an arbitrary cosmological constant represent the
analogs of the standard Schwarzschild-(A)dS solutions, which were
absent in the original Ho\v{r}ava model \cite{Park:2009zra}.

The fundamental constants of the speed of light $c$, Newton's
constant $G$, and the cosmological constant $\bar{\Lambda}$ are
defined as
\begin{equation}
c^2=\frac{\kappa^2\mu^2|\Lambda_W|}{8(3\lambda-1)^2}\,,\qquad
G=\frac{\kappa^2c^2}{16\pi(3\lambda-1)}\,,\qquad
\bar{\Lambda}=\frac{3}{2}\Lambda_W c^2.
\end{equation}

\subsection{Modified Friedman equations}

Consider the homogeneous and isotropic cosmological solution given
by the following metric
\begin{align}
  ds^2=-dt^2+a^2(t)\left[\frac{dr^2}{1-kr^2}+r^2\,(d\theta^2
   +\sin^2{\theta} \, d\phi ^2)\right] \,,
  \label{Einst:metric2}
\end{align}
where $k=+1,0,-1$ corresponds to a closed, flat, and open
universe, respectively.

We assume that the matter contribution takes the form of a perfect
fluid, with $\rho$ and $p$ the energy density and the pressure,
respectively, so that the modified Friedman equations in
Ho\v{r}ava gravity take the following form \cite{Park:2009zr}
\begin{eqnarray}
\left(\frac{\dot{a}}{a}\right)^2&=&\frac{\kappa^2}{6(3\lambda-1)}
\left[\rho + \epsilon
\frac{3\kappa^2\mu^2}{8(3\lambda-1)}\left(-\frac{k^2}{a^4}
+\frac{2k(\Lambda_W-\varpi)}{a^2}-\Lambda_W^2\right)\right]\,,
  \label{Hfieldeq1} \\
\frac{\ddot{a}}{a}&=&\frac{\kappa^2}{6(3\lambda-1)}
\left[-\frac{1}{2}(\rho+3p) + \epsilon
\frac{3\kappa^2\mu^2}{8(3\lambda-1)}\left(\frac{k^2}{a^4}
-\Lambda_W^2\right)\right]\,,
  \label{Hfieldeqs}
\end{eqnarray}
where $\epsilon = \pm 1$.

The analytic continuation $\mu^2 \rightarrow -\mu^2$ for the dS
case, i.e., $\Lambda_W>0$, is considered \cite{Park:2009zr}, and
the upper (lower) sign denotes the AdS (dS) case. It is
interesting to note that the higher derivative term appearing in
the action (\ref{Haction}) contributes with a $1/a^4$ term, and
only exists for $k \neq 0$. This term dominates for low values of
$a$, and the general relativistic cosmological solutions are
recovered for large scales.

\section{The Einstein static Universe in Ho\v{r}ava gravity
and perturbations}
\label{sec:III}

\subsection{Field equations}

For the Einstein static universe, $a=a_0={\rm const}$ and $k=1$,
the Ricci scalar becomes $R=6/a_0^2$, (note that for $\lambda=1$,
GR is obtained in the IR limit). Furthermore, we consider a linear
equation of state, $p=w \rho$, so that the field equations in this
case are expressed in the following manner
\begin{eqnarray}
  \rho_0 &=& \frac{\epsilon \kappa^2\mu^2}{a_0^4}
  \frac{\left[2+3a_0^2(1+w)\varpi \pm
  \sqrt{4-6a_0^2(1+w)(1+3w)\varpi}\right]}{6(1+w)^2(3\lambda-1)}\,,
  \label{Hfieldeqs2a}
   \\
  \Lambda_W a_0^2 &=& \frac{(1+3w)\mp\sqrt{4-6a_0^2
  (1+w)(1+3w)\varpi} }{3(1+w)}\,,
  \label{Hfieldeqs2}
\end{eqnarray}
where $\rho_0$ and $p_0$ are the unperturbed energy density and
isotropic pressure, respectively. Note that it is useful to
introduce the dimensionless parameters $\Omega:=a_0^2\varpi$ and
$\Lambda = \Lambda_W a_0^2$. These relationships are useful to be
written in this form, as one has a first glance at the existence
issue that $\rho_0$ and $\Lambda$ should be real, which imposes
the following condition
\begin{equation}
  2 - 3 (1+w)(1+3w) \Omega \geq 0\,.
\end{equation}
The above inequality needs to be analyzed for the three cases
$w<-1$, $-1<w<-1/3$ and $w>-1/3$ which places restrictions on the
allowed values of $\Omega$. However, as we are primarily
interested in a physically reasonable Einstein static universe, we
will furthermore require positivity of the energy density, $\rho_0
> 0$. To further simplify the subsequent analysis, we will also
assume $\lambda > 1/3$.

\begin{figure}[!ht]
\includegraphics[width=0.45\textwidth]{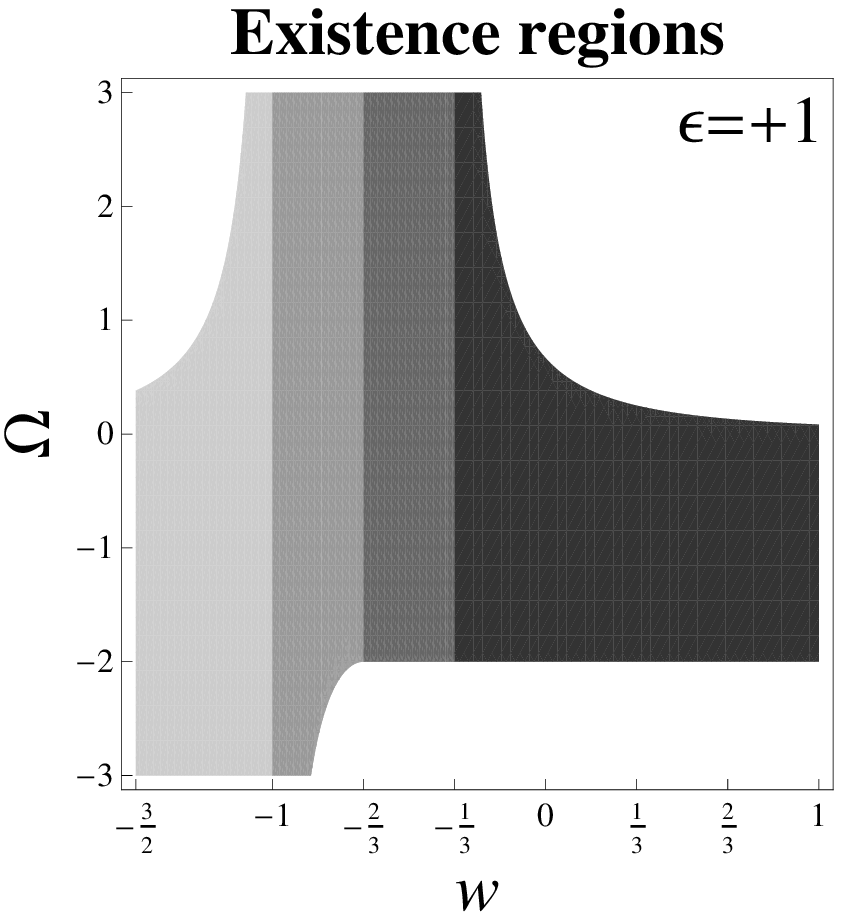}\hfill
\includegraphics[width=0.45\textwidth]{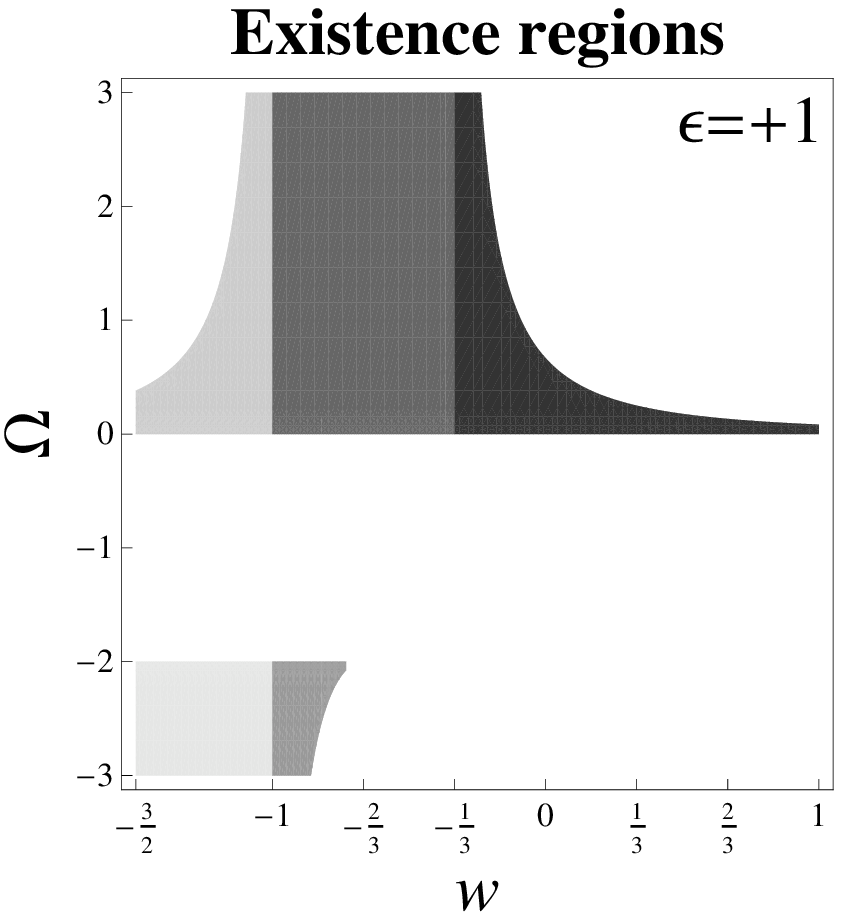}
\caption{Regions of existence in the $(w,\Omega)$ parameter space
for the specific case of $\epsilon=+1$. The left panel shows the
existence regions for the upper sign of Eq. (\ref{Hfieldeqs2a}),
while the right panel is for the lower sign. The different shades
of gray correspond to the different inequalitites.} \label{fig1}
\end{figure}

\begin{figure}[!ht]
\includegraphics[width=0.45\textwidth]{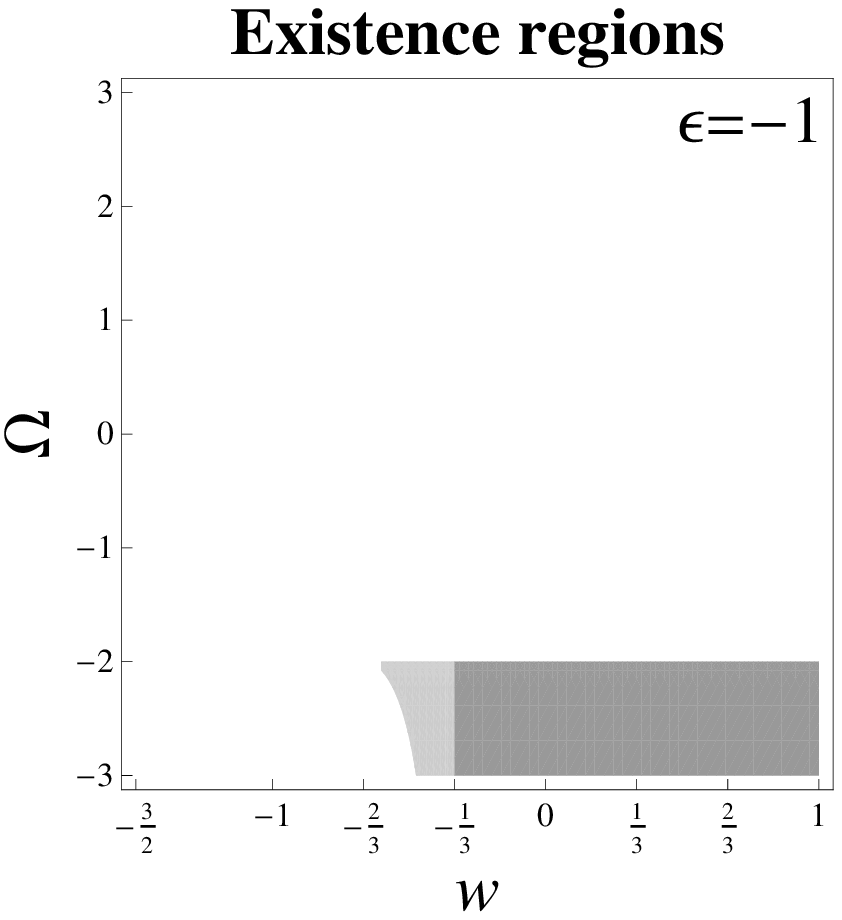}\hfill
\includegraphics[width=0.45\textwidth]{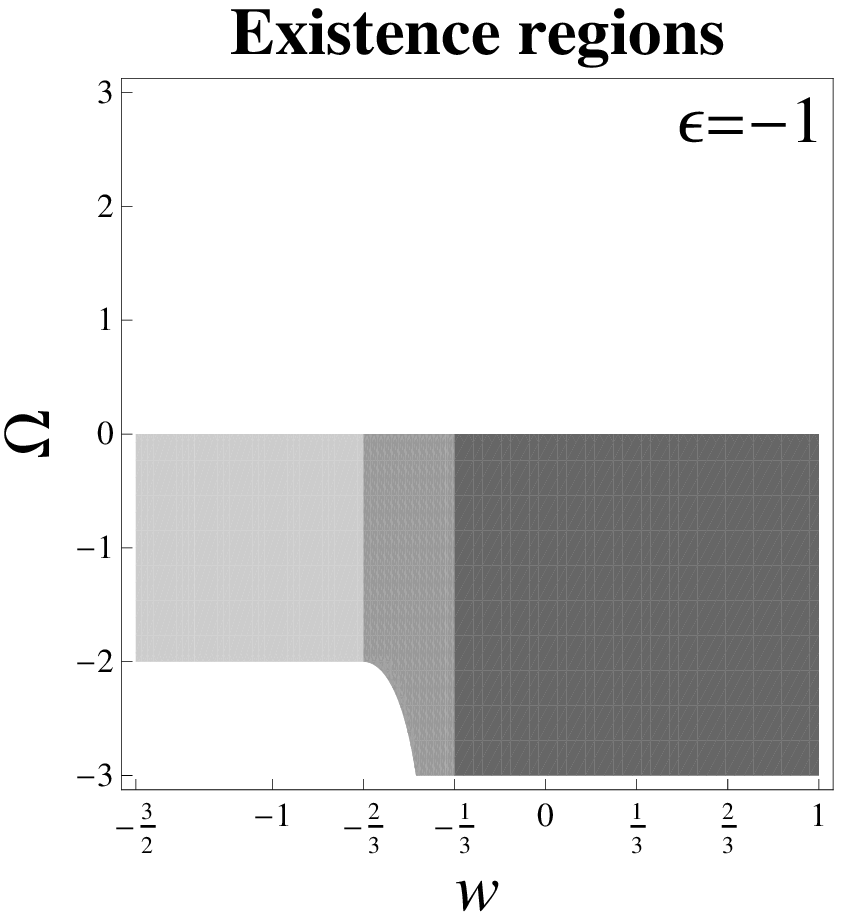}
\caption{Regions of existence in the $(w,\Omega)$ parameter space
for the specific case of $\epsilon=-1$. The left panel shows the
existence regions for the upper sign of Eq. (\ref{Hfieldeqs2a}),
while the right panel is for the lower sign. The different shades
of gray correspond to the different inequalitites.} \label{mfig1}
\end{figure}

These useful conditions imply the following existence conditions
for the upper sign of Eq. (\ref{Hfieldeqs2a}):
\begin{alignat}{3}
  &\epsilon = +1\,, &\qquad & w<-1\,, &\qquad & \Omega <
  \frac{2}{3}\frac{1}{(1+w)(1+3w)}\,,
  \nonumber \\
  &\epsilon = +1\,, &\qquad & -1<w<-2/3\,, &\qquad & \Omega >
  \frac{2}{3}\frac{1}{(1+w)(1+3w)}\,,
  \nonumber \\
  &\epsilon = +1\,, &\qquad & -2/3<w<-1/3\,, &\qquad & \Omega > -2\,,
  \nonumber\\
  &\epsilon = +1\,, &\qquad & -1/3<w\,, &\qquad & -2 < \Omega <
  \frac{2}{3}\frac{1}{(1+w)(1+3w)}\,,
  \nonumber \\
  &\epsilon = -1\,, &\qquad & -2/3<w<-1/3\,, &\qquad & \frac{2}{3}
  \frac{1}{(1+w)(1+3w)} < \Omega < -2\,,
  \nonumber \\
  &\epsilon = -1\,, &\qquad & -1/3<w\,, &\qquad & \Omega < -2\,,
  \label{exis_up}
\end{alignat}
which are depicted in the left plots of Figs. \ref{fig1} and
\ref{mfig1}, respectively.

For the lower sign,  of Eq. (\ref{Hfieldeqs2a}), we find:
\begin{alignat}{3}
  &\epsilon = +1\,, &\qquad & w<-1\,, &\qquad & \Omega < -2\,,
  \nonumber \\
  &\epsilon = +1\,, &\qquad & w<-1\,, &\qquad & 0 < \Omega <
  \frac{2}{3}\frac{1}{(1+w)(1+3w)}\,,
  \nonumber \\
  &\epsilon = +1\,, &\qquad & -1<w<-2/3\,, &\qquad &
  \frac{2}{3}\frac{1}{(1+w)(1+3w)} < \Omega < -2\,,
  \nonumber \\
  &\epsilon = +1\,, &\qquad & -1<w<-1/3\,, &\qquad & 0 < \Omega\,,
  \nonumber \\
  &\epsilon = +1\,, &\qquad & -1/3<w\,, &\qquad & 0 < \Omega
  < \frac{2}{3}\frac{1}{(1+w)(1+3w)}\,,
  \nonumber \\
  &\epsilon = -1\,, &\qquad & w<-2/3\,, &\qquad & -2 < \Omega < 0\,,
  \nonumber \\
  &\epsilon = -1\,, &\qquad & -2/3<w<-1/3\,, &\qquad & \frac{2}{3}
  \frac{1}{(1+w)(1+3w)} < \Omega < 0\,,
  \nonumber \\
  &\epsilon = -1\,, &\qquad & -1/3<w\,, &\qquad & \Omega < 0\,.
  \label{exis_dn}
\end{alignat}
which are depicted in the right plots of Figs. \ref{fig1} and
\ref{mfig1}, respectively.

The above inequalities with the upper and lower signs, and with
$\epsilon = +1$, are represented graphically in Fig.~\ref{fig1},
and with $\epsilon = -1$ in Fig.~\ref{mfig1}, respectively. Note
that there are many configurations which allow for an Einstein
static universe. This is in contrast with other modifications of
GR, such as $f(R)$ modified gravity, where there exists a unique
background Einstein static universe for every choice of $f(R)$.

\subsection{Linear homogeneous perturbations}

In what follows, we analyze the stability against linear
homogeneous perturbations around the Einstein static universe
given in Eqs.~(\ref{Hfieldeq1})-(\ref{Hfieldeqs}). Thus, we
introduce perturbations in the energy density and the metric scale
factor which only depend on time
\begin{align}
  \rho(t) = \rho_0+\delta\rho_1(t),
  \qquad
  a(t) = a_0+\delta a_1(t).
  \label{def:perturbations}
\end{align}

Now we consider adiabatic perturbations which also satisfy a
linear equation of state, $\delta p(t)=w\delta \rho(t)$ and
linearize the perturbed field equations. Firstly, we consider
Eq.~(\ref{Hfieldeq1}) which upon subtracting the background field
equation yields
\begin{equation}
  \delta \rho_1 = \frac{3\epsilon\kappa^2\mu^2}{2a_0^5(3\lambda-1)}
  (\Lambda-\Omega-1)\, \delta a_1\,.
\end{equation}
Next, we perturb the evolution equation~(\ref{Hfieldeqs}) and
eliminate the perturbed energy density by virtue of the latter
equation. The resulting second order differential equation for
$\delta a_1$ is given by
\begin{align}
  \delta a_1''(t) + \frac{\epsilon\kappa^4\mu^2}{8a_0^4(3\lambda-1)^2}
  \bigl[(1+\Lambda-\Omega)-3w(1-\Lambda+\Omega) \bigr]
  \delta a_1(t) = 0\,.
  \label{perturbeq2}
\end{align}
As only the sign of the prefactor of the second term is relevant,
we can rescale $\delta a_1(t)$ appropriately and consider
\begin{align}
  \delta a_1''(t) + \epsilon \bigl[(1+\Lambda-\Omega)
  +3w(-1+\Lambda-\Omega) \bigr] \delta a_1(t) = 0\,.
  \label{perturbeq}
\end{align}
Before solving this equation it should be noted that $\Lambda$ is
determined by the background, see Eq.~(\ref{Hfieldeqs2}) and hence
this quantity should also be substituted. We find
\begin{align}
  \Lambda-\Omega = \frac{(1-3\Omega)+3w(1-\Omega)\mp
  \sqrt{4-6(1+w)(1+3w)\Omega}}{3(1+w)}\,.
  \label{lambdabg}
\end{align}

Using the standard ansatz $a_1(t)=A\exp(iWt)$, where $A$ and $W$
are constants, we find that the above differential equation
(\ref{perturbeq}) provides the following solutions
\begin{align}
  W= \pm \sqrt{\epsilon}
  \sqrt{(1+\Lambda-\Omega)+3w(-1+\Lambda-\Omega)}\,.
  \label{sol:perturbeq}
\end{align}

\subsection{Stability regions}

To analyze the stability regions, first consider the cases
$\epsilon=+1$ ($\epsilon=-1$). Now, the stability conditions
impose that the factor within the square root of
Eq.~(\ref{sol:perturbeq}) is positive (negative), which is
translated by the following inequalities:
\begin{alignat}{2}
  & \epsilon = +1 & \qquad &
  (1+\Lambda-\Omega)+3w(-1+\Lambda-\Omega) > 0\,,\\
  & \epsilon = -1 & \qquad & (1+\Lambda-\Omega)
  +3w(-1+\Lambda-\Omega) < 0\,.
  \label{stability1}
\end{alignat}

\begin{figure}[!ht]
\includegraphics[width=0.45\textwidth]{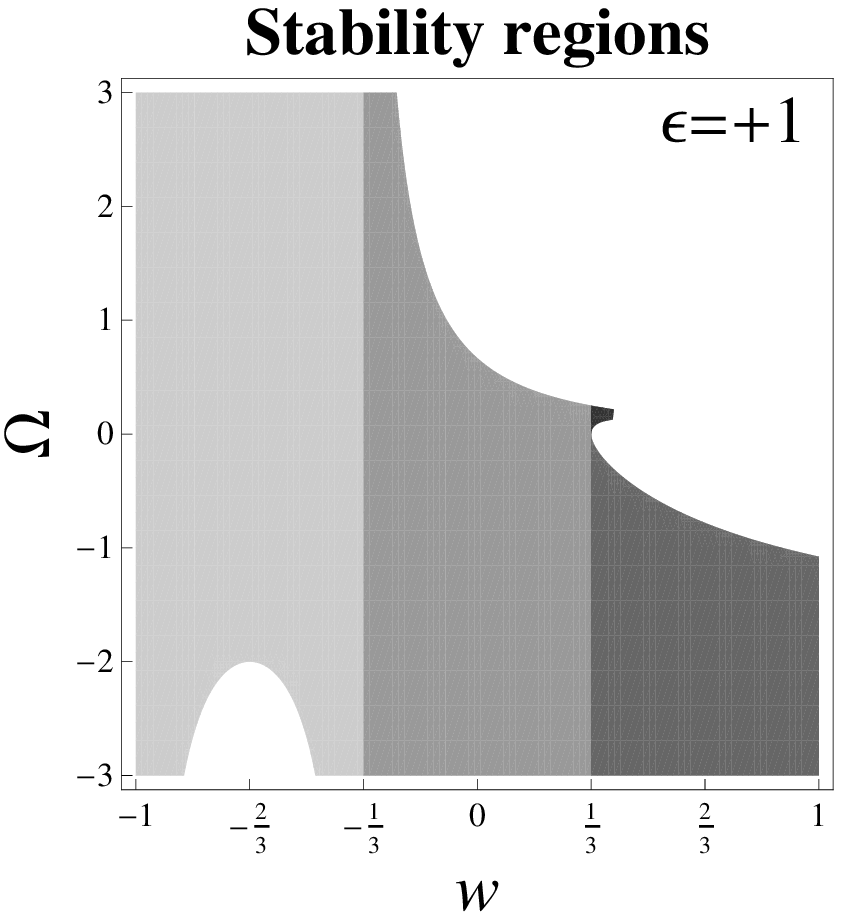}\hfill
\includegraphics[width=0.45\textwidth]{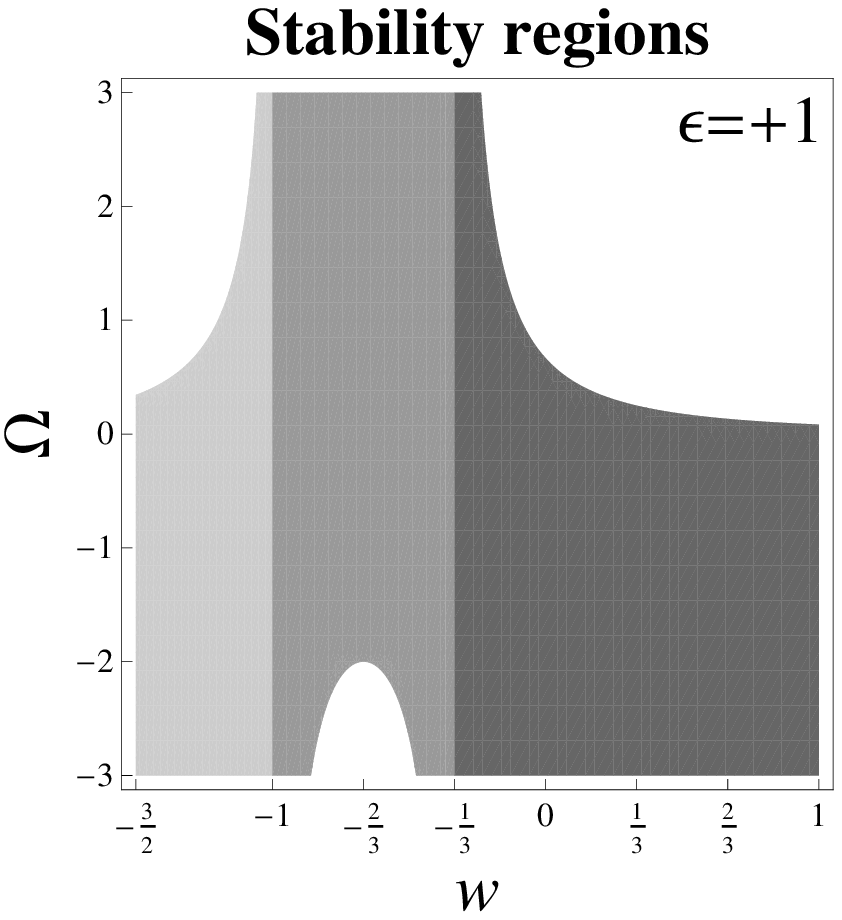}
\caption{Regions of stability in the $(w,\Omega)$ parameter space
(left panel, upper sign; right panel, lower sign), for the
specific case of $\epsilon=+1$. See the text for details.}
\label{fig2}
\end{figure}

\begin{figure}[!ht]
\includegraphics[width=0.45\textwidth]{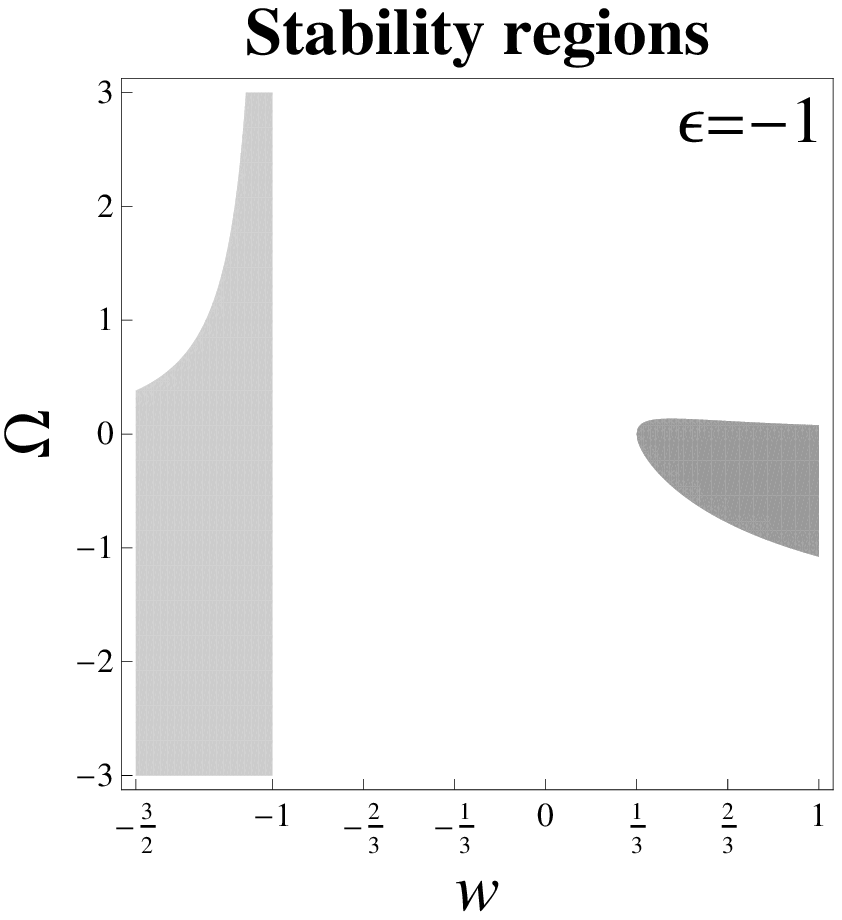}\hfill
\includegraphics[width=0.45\textwidth]{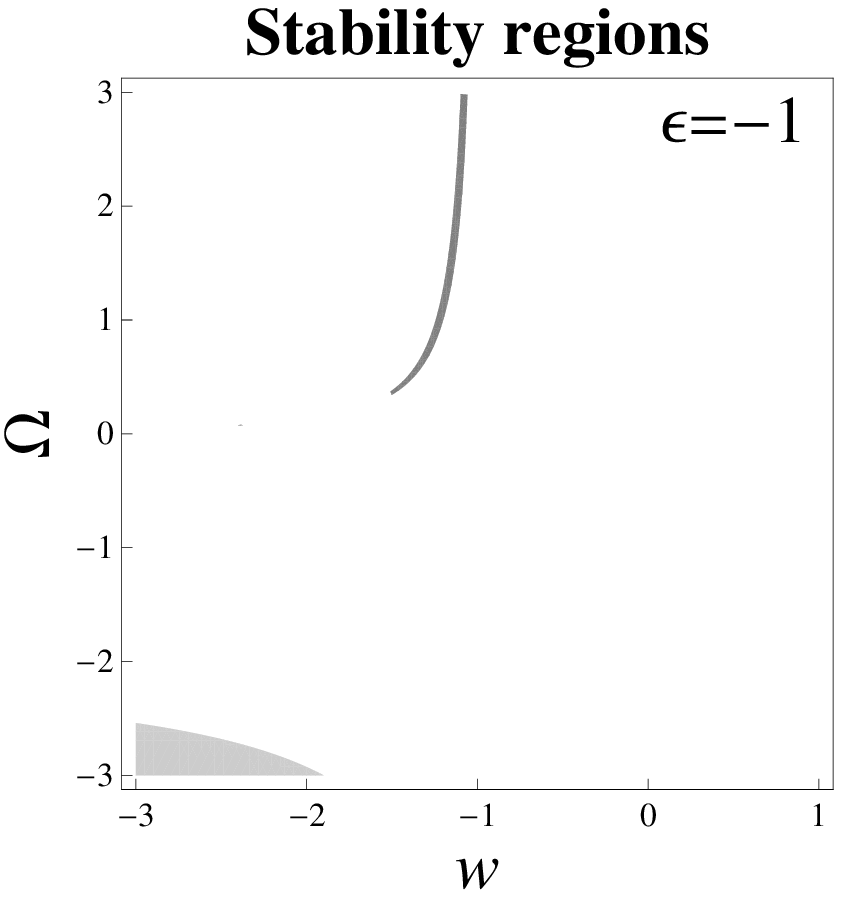}
\caption{Regions of stability in the $(w,\Omega)$ parameter space
(left panel, upper sign; right panel, lower sign), for the
specific case of $\epsilon=-1$. See the text for details.}
\label{mfig2}
\end{figure}

Since $\Lambda$ is determined by the background equations,
see~(\ref{lambdabg}), the equations become more complicated. For
the upper sign we have:
\begin{alignat}{3}
  &\epsilon = +1\,, &\qquad & -1<w<-1/3\,, & \qquad & \Omega >
  \frac{2}{3}\frac{1}{(1+w)(1+3w)}\,,
  \nonumber \\
  &\epsilon = +1\,, &\qquad & -1/3<w<1/3\,, & \qquad & \Omega <
  \frac{2}{3}\frac{1}{(1+w)(1+3w)}\,,
  \nonumber \\
  &\epsilon = +1\,, &\qquad & 1/3<w\,, &\qquad & \Omega <
  \frac{1-3w}{1+3w}-\frac{1}{\sqrt{3}}\sqrt{\frac{-1+3w}{1+w}}\,,
  \nonumber \\
  &\epsilon = +1\,, &\qquad & 1/3<w\,, &\qquad &
  \frac{1-3w}{1+3w}+\frac{1}{\sqrt{3}}\sqrt{\frac{-1+3w}{1+w}} <
  \Omega < \frac{2}{3}\frac{1}{(1+w)(1+3w)}\,,
  \nonumber \\
  &\epsilon = -1\,, &\qquad & w<-1\,, &\qquad & \Omega <
  \frac{2}{3}\frac{1}{(1+w)(1+3w)}\,,
  \nonumber \\
  &\epsilon = -1\,, &\qquad & 1/3<w\,, &\qquad & \frac{1-3w}{1+3w}
  -\frac{1}{\sqrt{3}}\sqrt{\frac{-1+3w}{1+w}} < \Omega <
  \frac{1-3w}{1+3w}+\frac{1}{\sqrt{3}}\sqrt{\frac{-1+3w}{1+w}}\,,
  \label{stab_up}
\end{alignat}
which are depicted in the left plots of Figs.~\ref{fig2} and
\ref{mfig2}, respectively.

Finally, for the lower sign we obtain:
\begin{alignat}{3}
  &\epsilon = +1\,, &\qquad & w<-1\,, & \qquad & \frac{1-3w}{1+3w}
  -\frac{1}{\sqrt{3}}\sqrt{\frac{-1+3w}{1+w}} < \Omega <
  \frac{1-3w}{1+3w}+\frac{1}{\sqrt{3}}\sqrt{\frac{-1+3w}{1+w}}\,,
  \nonumber \\
  &\epsilon = +1\,, &\qquad & -1<w<-1/3\,, & \qquad & \Omega >
  \frac{2}{3}\frac{1}{(1+w)(1+3w)}\,,
  \nonumber \\
  &\epsilon = +1\,, &\qquad & -1/3<w\,, & \qquad & \Omega <
  \frac{2}{3}\frac{1}{(1+w)(1+3w)}\,,
  \nonumber \\
  &\epsilon = -1\,, &\qquad & w<-1\,, & \qquad & \Omega <
  \frac{1-3w}{1+3w}-\frac{1}{\sqrt{3}}\sqrt{\frac{-1+3w}{1+w}}\,,
  \nonumber \\
  &\epsilon = -1\,, &\qquad & w<-1\,, &\qquad &
  \frac{1-3w}{1+3w}+\frac{1}{\sqrt{3}}\sqrt{\frac{-1+3w}{1+w}} <
  \Omega < \frac{2}{3}\frac{1}{(1+w)(1+3w)}\,,
  \label{stab_dn}
\end{alignat}
which are depicted in the right plots of Figs.~\ref{fig2} and
\ref{mfig2}, respectively.

\begin{figure}[!ht]
\includegraphics[width=0.45\textwidth]{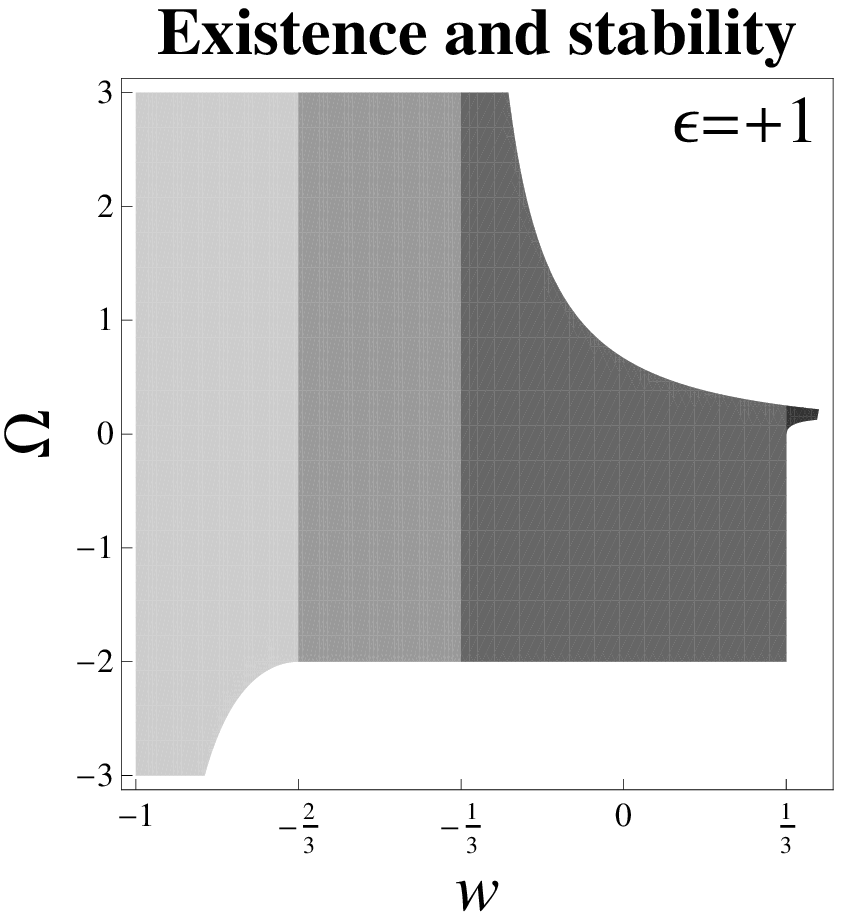}\hfill
\includegraphics[width=0.45\textwidth]{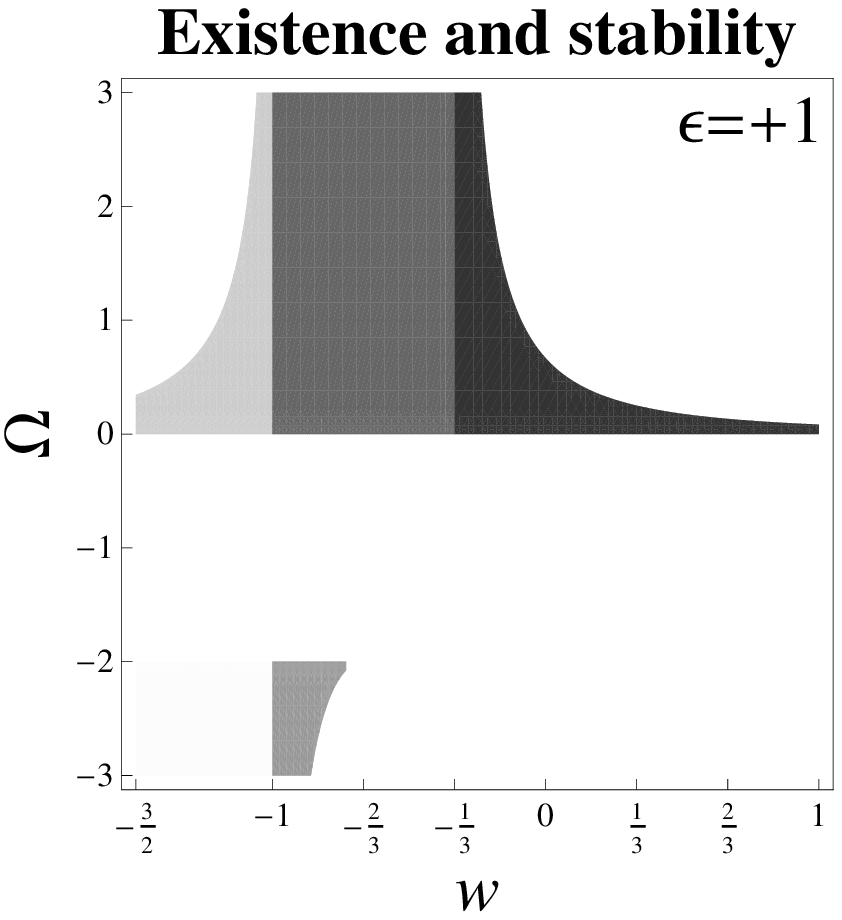}
\caption{Combined region of existence and stability in the
$(w,\Omega)$ parameter space (left panel, upper sign; right panel,
lower sign) for the specific case of $\epsilon=+1$.} \label{fig3}
\end{figure}

By combining the existence conditions with the stability
conditions, we can identify a large class of Einstein static
universes which are stable with respect to homogeneous
perturbations. It should also be noted that for the upper sign a
Einstein static universe of phantom matter ($w<-1$) exists.
Superimposing the inequality plots Figs.~\ref{fig1}--\ref{mfig2}
we can picture the complete parameter space for which the Einstein
static universe in IR modified Ho\v{r}ava gravity exits and is
stable with respect to linear homogeneous perturbations. This is
depicted in Fig.~\ref{fig3} for the specific case of
$\epsilon=+1$. For the specific case of $\epsilon=-1$, we verify
the nonexistence of the Einstein static universe
existence/stability regions.

For every parameter choice we can compute the actual value of
$\Lambda$ by using Eq.~(\ref{Hfieldeqs2}). Note, however, that 
it is very involved to represent the values $\Lambda$ can attain 
in general because this would require to plot a surface in the 
$(w,\Omega,\Lambda)$ parameter space which satisfies various 
inequalities simultaneously.

\section{Summary and discussion}
\label{sec:concl}

The Einstein static universe has recently been revived as the
asymptotic origin of an emergent universe, namely, as an
inflationary cosmology without a singularity \cite{Ellis:2002we}.
The role of positive curvature, negligible at late times, is
crucial in the early universe, as it allows these cosmologies to
inflate and later reheat to a hot big-bang epoch. An attractive
feature of these cosmological models is the absence of a
singularity, of an `initial time', of the horizon problem, and the
quantum regime can even be avoided. Furthermore, the Einstein
static universe was found to be neutrally stable against
inhomogeneous linear vector and tensor perturbations, and against
scalar density perturbations provided that the speed of sound
satisfies $c_{\rm s}^2>1/5$ \cite{Barrow:2003ni}. Further issues
related to the stability of the Einstein static universe may be
found in Ref.~\cite{Barrow}.

In this work we have analyzed linear homogeneous perturbations
around the Einstein static universe in the context of IR modified
Ho\v{r}ava gravity. In particular, perturbations in the energy
density and the metric scale factor were introduced, a linear
equation of state, $p(t)=w\rho(t)$, was considered, and finally
the linearized perturbed field equations and the dynamics of the
solutions were analyzed. It was shown that stable modes for all
equation of state parameters $w$ exist and, in particular, the
complete parameter space for which the Einstein static universe in
IR modified Ho\v{r}ava gravity exits and is stable with respect to
linear homogeneous perturbations was presented. Thus, as in
Refs.~\cite{Boehmer:2007tr,Goswami:2008fs,Seahra:2009ft,Bohmer:2009fc}
our results show that perturbation theory of modified theories of
gravity present a richer stability/instability structure that in
general relativity. Finally, it is of interest to extend our
results to inhomogeneous perturbations in the spirit of
Ref.~\cite{Seahra:2009ft}, and to include the canonical scalar
field case. Work along these lines is presently underway.


\end{document}